\documentclass[12pt]{article}
\usepackage{graphicx}
\advance\textwidth by 3truecm
\advance\oddsidemargin by -1.5truecm

\newcommand{\kB}{k_{\mathrm{B}}}
\newcommand{\pg}{{\mathrm{pg}}}

\begin{document}
\title{\normalsize\bf LETTER\\[\baselineskip]
Scaling of the superconducting transition temperature\\
in underdoped high-$T_c$ cuprates with a pseudogap energy:\\
Does this support the anyon model of their superfluidity?}
\author{\normalsize G. G. N. Angilella$^1$, N. H. March$^{2,3}$,
   R. Pucci$^1$\\
\normalsize\emph{$^1$Dipartimento di Fisica e Astronomia, Universit\`a di
   Catania,}\\
\normalsize\emph{and Lab. MATIS-INFM, and CNISM, Sez. Catania, and INFN, Sez. Catania,}\\
\normalsize\emph{Via S. Sofia, 64, I-95123 Catania, Italy}\\
\normalsize\emph{$^2$Oxford University, Oxford, UK}\\
\normalsize\emph{$^3$Department of Physics, University of Antwerp,}\\ 
\normalsize\emph{Groenenborgerlaan 171, B-2020 Antwerp, Belgium}}

\date{}

\maketitle

\abstract{%
In earlier work, we have been concerned with the scaling properties of some
classes of superconductors, specifically with heavy Fermion materials and with
five bcc transition metals of BCS character.
Both of these classes of superconductors were three-dimensional but here we are
concerned solely with quasi-two-dimensional high-$T_c$ cuprates in the
underdoped region of their phase diagram.
A characteristic feature of this part of the phase diagram is the existence of a
pseudogap (pg).
We therefore build our approach around the assumption that $\kB T_c / E_\pg$ is
the basic dimensionless ratio on which to focus, where the energy $E_\pg$
introduced above is a measure of the pseudogap.

Since anyon fractional statistics apply to two-dimensional assemblies, we expect
the fractional statistics parameter allowing `interpolation' between Fermi-Dirac
and Bose-Einstein statistical distribution functions as limiting cases to play a
significant role in determining $\kB T_c / E_\pg$ and experimental data are
analyzed with this in mind.
\\[0.5\baselineskip]
}

\bigskip

In recent work
\cite{Angilella:00b,Angilella:01a,Angilella:03d,Angilella:04e,Angilella:04l},
we have been concerned with the connection between superconducting transition
temperature $T_c$ and other physical properties in (a) heavy Fermion materials
\cite{Angilella:00b,Angilella:01a,Angilella:04l} and (b) five body-centred
cubic (bcc) transition metals \cite{Angilella:04e}. For case (a), $\kB T_c$ was
shown to correlate strongly with a `kinetic energy of localization' $\hbar^2
/m^\ast \xi^2$, where $m^\ast$ is the effective mass of the charge carriers,
while $\xi$ is the coherence length. In contrast, in case (b) an intimate
relationship between $T_c$ and elastic constants was displayed, showing very
directly thereby that these bcc transition metals were BCS-like
superconductors.

Here, we restrict ourselves solely to the quasi two-dimensional high-$T_c$
cuprates.
Because it is known that these materials in the underdoped region of the phase
diagram are associated with a pseudogap, it seemed to us natural to consider the
ratio $T_c /T^\ast$, where $\kB T^\ast$ is a measure of the energy of the
pseudogap.
Thus, bearing in mind that anyon fractional statistics \cite{Wilczek:90} are
associated with two dimensions, we propose that
\begin{equation}
\frac{T_c}{T^\ast} = F(T^\ast , \alpha),
\label{eq:correlation}
\end{equation}
where the, as yet unknown, function $F$ depends on $T^\ast$ itself, and on the
anyon fractional statistics parameter $\alpha$ chosen to lie in the range
between 0 and 1.
In addition, we must expect that, whereas in BCS theory $T^\ast \to
\Theta_{\mathrm{D}}$, where $\Theta_{\mathrm{D}}$ is the Debye temperature, and
no anyons exist in three dimensions, there will be some coupling parameter as
for the strength of the electron-phonon interaction in BCS theory.
While we anticipate that Eq.~(\ref{eq:correlation}) should obtain in the
quasi-2D high-$T_c$ cuprates independent of any particular simplifying model, we
intend below to exemplify our assumptions embodied in Eq.~(\ref{eq:correlation})
by appealing to a very recent and specific 2D model \cite{Fine:05,Fine:04},
which we propose to generalize by heuristic arguments to embrace the fractional
statistics parameter $\alpha$.

The mean-field $T_c$ has been correlated to the quantity $E_\pg$ related to a
measure of the
pseudogap by Fine \cite{Fine:05} within a model \cite{Fine:04} based on the
existence of stripes in the CuO$_2$ planes of high-$T_c$ superconductors. The
resulting correlation can be presented as \cite{Fine:05}
\begin{equation}
\kB T_c = \frac{g^2}{8E_\pg} \frac{e^{E_\pg /\kB T_c}
-1}{e^{E_\pg/\kB T_c} +1} ,
\label{eq:Fine}
\end{equation}
where $g$ is the coupling constant between stripes and the antiferromagnetic
(AFM) domains between the stripes.

We now turn to the heuristic generalization of Fine's model for $T_c$ displayed
in Eq.~(\ref{eq:Fine}).
We appeal first to the simple collision model used by one of us \cite{March:93b}
(see also Ref.~\cite{March:97b}), which allowed a partial unification of Fermi-Dirac
(FD), Bose-Einstein (BE), and anyon fractional statistics.
If the statistical distribution function in each case is denoted by
$f(\epsilon)$, with $\epsilon$ the particle energy, then the unifying equation
in Refs.~\cite{March:93b,March:97b} was
\begin{equation}
\frac{1}{f(\epsilon)} = \exp \left(\frac{\epsilon-\mu}{\kB T} \right) + a,
\end{equation}
where $a$ was assumed to depend only on $\alpha$.
Then the choice $a=2\alpha-1$ gave $a$ correctly for BE statistics with
$\alpha=0$ and $a=-1$, and for FD statistics with $\alpha=1$ and $a=1$.
The subsequent microscopic theory of Wu \cite{Wu:94} showed that $a$ away from
these endpoints $\alpha=0$ and 1 also depended on $\epsilon/\kB T$.
Wu's result takes the form
\begin{equation}
n(\epsilon-\mu) = \frac{1}{w[e^{(\epsilon-\mu)/\kB T}] + \alpha} ,
\label{eq:Wu1}
\end{equation}
where $\epsilon$ is the energy level, $\mu$ the chemical potential, and $\alpha$
is the fractional statistics parameter introduced above, ranging between the limiting values $\alpha=0$ and $\alpha=1$,
for which Eq.~(\ref{eq:Wu1}) reduces to the familiar Bose-Einstein and
Fermi-Dirac
distributions, respectively.
The case $\alpha=\frac{1}{2}$ refers to `semions'.
In Eq.~(\ref{eq:Wu1}), the `generalized exponential' $w(\zeta)$ obeys the
functional equation \cite{Wu:94}
\begin{equation}
w^\alpha (\zeta) [1+w(\zeta)]^{1-\alpha} = \zeta \equiv e^{(\epsilon-\mu)/\kB T}
.
\label{eq:Wu2}
\end{equation}

We then propose to generalize Fine's equation~(\ref{eq:Fine}) above to embrace the
cases of anyon statistics as follows:
\begin{equation}
\frac{\kB T_c}{E_\pg} = \frac{g^2}{8E_\pg^2} 
\frac{w(e^{E_\pg/\kB T_c} ) + \alpha}{2 e^{E_\pg/\kB T_c} - w (e^{E_\pg/\kB T_c}
) - \alpha} ,
\label{eq:Fine1}
\end{equation}
where we have divided both sides of Eq.~(\ref{eq:Fine}) by $E_\pg$, in order to
form the dimensionless ratios $\kB T_c / E_\pg$ and $8E_\pg^2 /g^2$.
It may be checked that Eq.~(\ref{eq:Fine1}) reduces to Fine's formula,
Eq.~(\ref{eq:Fine}), in the limit $\alpha=0$.

\begin{figure}[t]
\centering
\includegraphics[height=0.7\columnwidth,angle=-90]{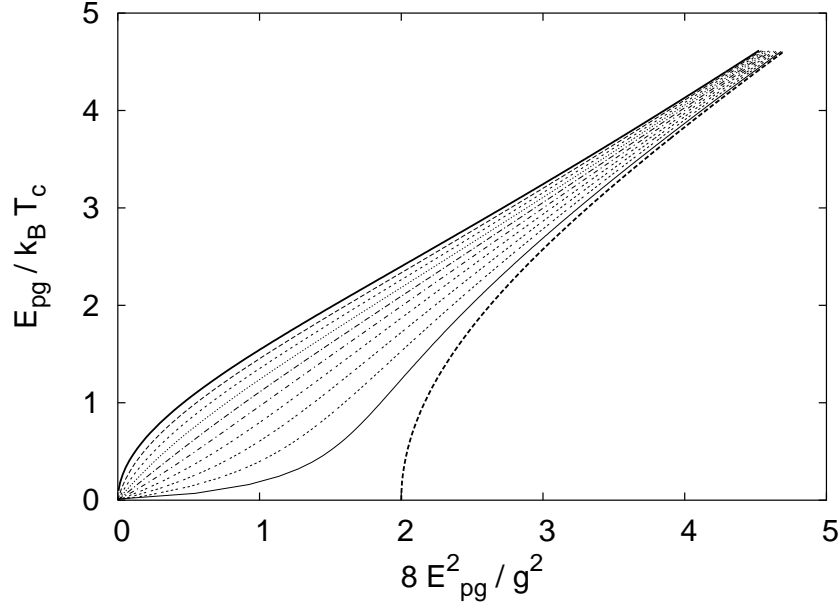}
\caption{Showing our proposed generalization of Fine's formula,
Eq.~(\ref{eq:Fine1}), to the case of anyon statistics.
The anyon parameter ranges from $\alpha=0$ (top curve, solid line) to $\alpha=1$
(bottom curve, dashed line).}
\label{fig:Wu}
\end{figure}

Fig.~\ref{fig:Wu} shows the numerical solution of Eq.~(\ref{eq:Fine1}) for the
inverse ratio $E_\pg /\kB T_c$ plotted as a function of the dimensionless
variable $8 E^2_\pg /g^2$, with $g$ denoting, as mentioned above, Fine's
coupling constant.
The different curves are characterized by the specific values of the parameter
$\alpha$ recorded in Fig.~\ref{fig:Wu}.
Above the value $E_\pg /\kB T_c \simeq 1$, these curves are (i) rather linear
and (ii) have slopes which vary only weakly with $\alpha$ until it reaches
around the semion value $\alpha=\frac{1}{2}$.
We have also plotted, for completeness of the consequences of
Eq.~(\ref{eq:Fine1}), the curve for $\alpha=1$ though we do not anticipate it
will have significance for the high-$T_c$ cuprates, as the ensuing discussion
will indicate.

\begin{figure}[t]
\centering
\includegraphics[height=0.7\columnwidth,angle=-90]{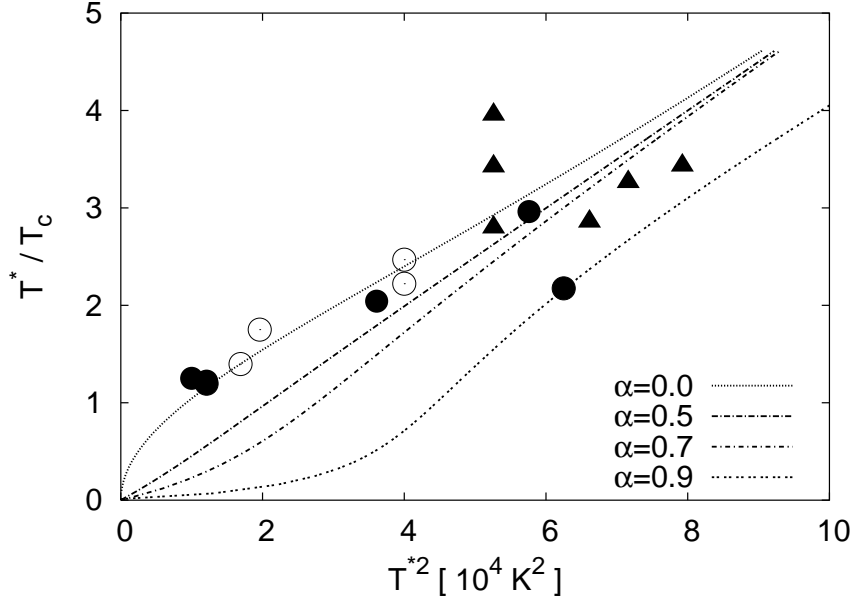}
\caption{Ratio of pseudogap and critical temperature, $T^\ast /T_c$, versus square
of pseudogap temperature, $T^{\ast2}$, for several high-$T_c$ superconductors in
the pseudogap regime (data taken from Tables~1 and 2 of Ref.~\cite{Timusk:99}).
Filled and open circles refer to pseudogap temperatures $T^\ast$ from Knight
shift and NMR relaxation rates measurements, respectively, while filled
triangles refer to maximum pseudogap, as derived from in-plane optical
scattering measurements \cite{Timusk:99}.
Lines are instances of Eq.~(\ref{eq:Fine1}), with 
$\kB T^\ast \equiv E_\pg$, and $g^2 /8=2 \cdot 10^4$~K$^2$ for the cases
$\alpha=0,\, 0.5,\, 0.7$ cases, and $g^2 /8= 2.4 \cdot 10^4$~K$^2$ for the case $\alpha=0.9$.}
\label{fig:Timusk}
\end{figure}

This is the point at which we invoke explicit experimental data for the
high-$T_c$ cuprates.
The review of Timusk and Statt~\cite{Timusk:99} compares $T_c$ with $T^\ast$
determined by Knight shift and by NMR relaxation experiments in their Table~1,
as well as by in-plane optical scattering measurements in their Table~2.
In Fig.~\ref{fig:Timusk},
and in the absence of knowledge of what to take for
the coupling constant $g$, we assume as a starting point that it is material
independent and plot therefore, but now from experiment, $T^\ast / T_c$ versus
$T^{\ast 2}$ (cf. Fig.~\ref{fig:Wu}).
There is already clear evidence for the shape displayed in Fig.~\ref{fig:Wu}.

For most of the points in Fig.~\ref{fig:Timusk}, $\alpha$ small is the
appropriate choice.
But for the highest $T_c$ material HgBa$_2$Ca$_2$Cu$_3$O$_{8+\delta}$ with $T_c
= 115$~K and $T^\ast = 250$~K \cite{Timusk:99}, we can fit this rather isolated
point by taking $\alpha=0.9$ and making a change in coupling strength.
It is tempting therefore to believe that the highest $T_c$ cuprates with $T_c
\sim 140$~K might correspond also to a large fractional statistics parameter
$\alpha$.

To summarize, we have been able to give some substantial support to our basic
assumption Eq.~(\ref{eq:correlation}) by making direct use of experimental data
on the pseudogap.
By heuristic generalization of Fine's mean field solution of his 2D stripe
model, we have exhibited how we expect the fractional statistics parameter
$\alpha$ to influence the behaviour of $T^\ast / T_c$.
Present indications are that higher $T_c$ materials should correspond to larger
$\alpha$, but the material dependence of the basic coupling strength ($g$, in
Fine's model) remains to be studied more fundamentally.

\subsection*{Acknowledgements}

One of us (NHM) acknowledges that his contribution to this Letter was made
during a stay in the Department of Physics and Astronomy, University of Catania.
He thanks Professors F. Catara, R. Pucci, and E. Rimini for generous
hospitality. NHM wishes also to acknowledge travel and maintenance support from the University of
Antwerp, and to thank especially Professor D. Van~Dyck for his continuing
hospitality.

\bibliographystyle{pcl}
\bibliography{a,b,c,d,e,f,g,h,i,j,k,l,m,n,o,p,q,r,s,t,u,v,w,x,y,z,zzproceedings,Angilella}

\end{document}